\documentclass[journal]{IEEEtran}

\IEEEoverridecommandlockouts

\usepackage{amsmath,amssymb,amsfonts}
\usepackage{graphicx}
\usepackage{textcomp}

\usepackage[nocompress]{cite}

\usepackage{rotating}
\usepackage{multirow}
\usepackage{array}
\usepackage{xcolor}
\usepackage{url}
\usepackage{pgfplots}
\usepackage{verbatim}
\usepackage{tikz}
\usepackage{algorithm}
\usepackage{algpseudocode}
\usepackage{breqn}
\usepackage{amsmath, amsfonts}
\usepackage{amsmath, amssymb}

\usepackage{soul}

\usepackage[utf8]{inputenc}
\usepackage{nomencl}
\makenomenclature

\usetikzlibrary{mindmap,shadows,backgrounds}
\graphicspath{{figures/}}
\pgfplotsset{compat=1.14}

\def\BibTeX{{\rm B\kern-.05em{\sc i\kern-.025em b}\kern-.08em
    T\kern-.1667em\lower.7ex\hbox{E}\kern-.125emX}}

\ifCLASSINFOpdf

\else

\fi

\begin{document}

\title{AI-enabled Future Wireless Networks: Challenges, Opportunities and Open Issues}

\author{\IEEEauthorblockN{Medhat Elsayed and Melike Erol-Kantarci}
\\
\IEEEauthorblockA{\textit{School of Electrical Engineering and Computer Science} \\
\textit{University of Ottawa}\\
Ottawa, Canada \\
Emails: \{melsa034,melike.erolkantarci\}@uottawa.ca}}

\maketitle

\begin{abstract}
A plethora of demanding services and use cases mandate a revolutionary shift in the management of future wireless network resources. Indeed, when tight quality of service demands of applications are combined with increased complexity of the network, legacy network management routines will become unfeasible in 6G.  Artificial Intelligence (AI) is emerging as a fundamental enabler to orchestrate the network resources from bottom to top. AI-enabled radio access and AI-enabled core will open up new opportunities for automated configuration of 6G.  On the other hand, there are many challenges in AI-enabled networks that need to be addressed. Long convergence time, memory complexity, and complex behaviour of machine learning algorithms under uncertainty as well as highly dynamic channel, traffic and mobility conditions of the network contribute to the challenges. In this paper, we survey the state-of-art research in utilizing machine learning techniques in improving the performance of wireless networks. In addition, we identify challenges and open issues to provide a roadmap for the researchers. 
\end{abstract}

\IEEEpeerreviewmaketitle

\section{Introduction}
\label{sec:Intro}

\begin{figure*}
	\centering
		\includegraphics[scale = 0.35]{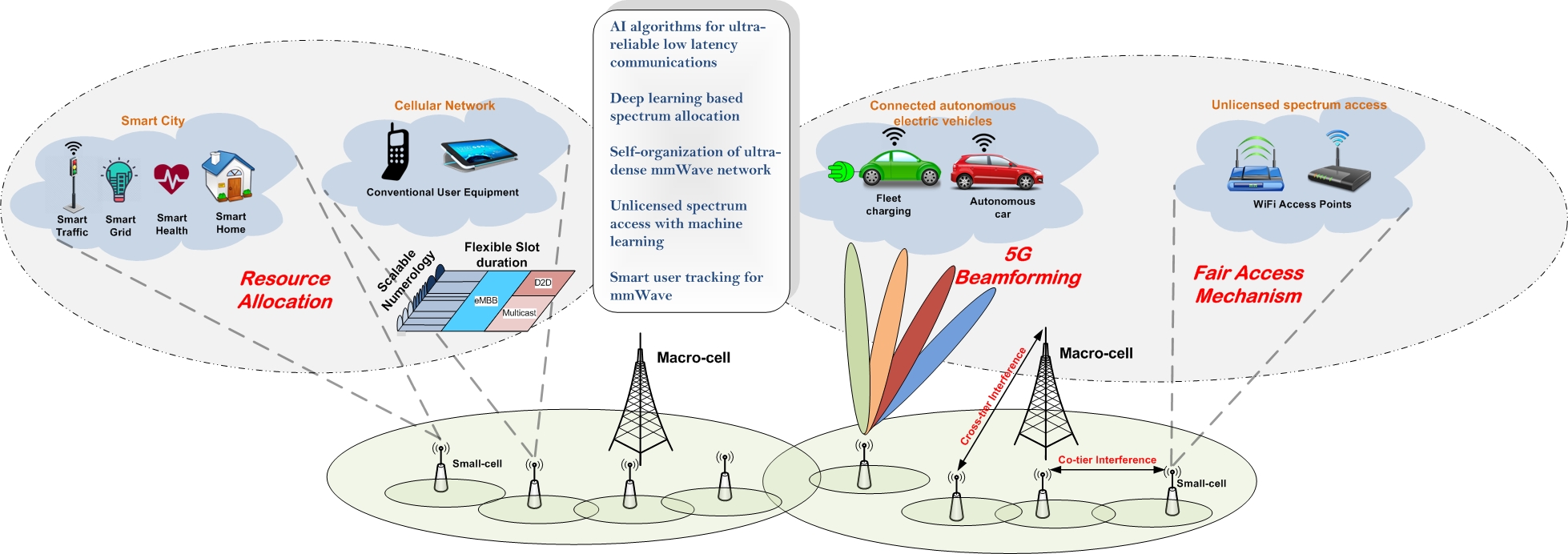}
	\caption{AI-enabled future wireless network and services.}
	\label{fig:5G}
\end{figure*}

 Future wireless networks are expected to support a multitude of services. According to the International Telecommunication Union, 5G network services can be classified into three service scenarios: Enhanced Mobile Broadband (eMBB), Ultra-Reliable and Low-latency Communications (uRLLC), and massive Machine Type Communications (mMTC) \cite{ITUR}. Heterogeneous devices of different quality of service demands will require intelligent and flexible allocation of network resources in response to network dynamics. For instance, a highly reliable and low-latency network is needed to enable rapid transfer of messages between connected autonomous vehicles. At the same time, the same physical infrastructure is expected to serve users with high-quality video demand or even mobile Augmented/Virtual Reality entertainment applications. Next-generation wireless networks, i.e. 5G and the upcoming 6G, are expected to accommodate diverse use cases. In particular, the heterogeneous traffic coming from mobile, vehicular, smart grid and tactile domains, calls for efficient utilization of network resources to maintain quality of service demands of each application. In addition, resource efficiency, reliability, and robustness are becoming more stringent for 5G and beyond networks. To meet this, 6G networks must incorporate a paradigm shift in network resource optimization, in which efficient and intelligent resource management techniques have to be employed. Artificial intelligence, or more specifically machine learning algorithms stand as promising tools to intelligently manage the networks such that network efficiency, reliability, robustness goals are achieved and quality of service demands are satisfied. The opportunities that arise from learning the environment parameters under varying behavior of the wireless channel, positions AI-enabled 5G and 6G  superior to preceding generations of wireless networks. Fig. \ref{fig:5G} highlights some wireless problems and applications that can leverage the potential of artificial intelligence.\\

In the literature, several research efforts have addressed radio resource allocation, Device-to-Device communications, access to unlicensed bands, routing, security, and fault management using machine learning \cite{8255757}. Among these, radio resource allocation has seen a significant interest from the research community. In particular, distributed algorithms have been proposed to enable each base station to learn radio resource allocation parameters independently, which facilitates the deployment of small base stations readily without the need of pre-configuration. Among a wide variety of ML algorithms, deep learning has been employed widely which are extensively surveyed in [3]. Among a wide variety of machine learning algorithms, deep learning has been employed widely in many works. A comprehensive survey of the applications of deep learning algorithms to different network layers can be found in \cite{8382166}. Furthermore, \cite{Hanzo} discusses various machine learning techniques and provides visionary use cases for their potential application to Multiple Input and Mulitple Output (MIMO), heterogeneous networks and cognitive radio. \\

Unlike previous surveys, in this paper, we provide the state-of-the-art in AI-enabled techniques with a focus on resource allocation, spectrum access, base station deployment and energy efficiency in wide variety of uses cases, including mobile broadband, Tactile Internet, Device-to-device communications and unmanned aerial vehicles. In addition, we present a deep reinforcement learning based solution to resource allocation which provides significant improvement in delay performance. We also discuss challenges and open issues related to AI-enabled future wireless networks. \\

The paper is organized as follows. Section \ref{sec:ML} presents a background on machine learning methods. In Section \ref{sec:RA}, we briefly explain the resource allocation problem, followed by Section \ref{sec:RAML} where we present the state-of-art work in wireless resource allocation using machine learning. In section \ref{sec:others} we discuss recent works that address deployment, spectrum access and energy-efficiency using machine learning. In section \ref{sec:DMDQ}, we discuss our recent work on resource allocation using Deep Q-learning. Finally, section \ref{sec:conc} concludes the paper by presenting challenges and open issues in the field. 

\section{Background on Machine Learning}
\label{sec:ML}

Over the past decade, the huge growth in data across many different fields resulted in big data challenge which amplified the need for intelligent data analysis schemes. Various machine learning methods emerged, such as deep learning, and they have been used along with traditional machine learning methods to cope with the big data problem. Recently they have been adopted in wireless networks. Therefore, in this section we give a brief overview of widely used techniques. \\

Machine learning schemes can be classified into four main categories: Supervised learning, unsupervised learning, semi-supervised learning, and reinforcement learning. These four categories differ in the way the algorithm is being trained \cite{Alpaydin}. In supervised learning, the training is performed initially by some labeled data. The labeled data represents a set of inputs with their corresponding outputs, known beforehand. Therefore, supervised learning algorithms are well-suited to applications with historical data. Feature extraction and classification has been applied to several signal processing problems. In classification, the task is to identify which set of categories a new observation belongs to. In contrast, unsupervised learning algorithms aim to infer features in the data, thus inferring the implied structure. semi-supervised learning algorithms use both labeled and unlabeled data. Finally, reinforcement learning uses data from the implementation instead of historical data. The aim of reinforcement learning is to improve the performance of an agent in a certain task using feedback from the environment. As such, the agent's goal is to predict the next action to take to earn the biggest final reward. Reinforcement learning is unsupervised however the way of learning is different than other unsupervised learning techniques. Rather than learning the structure of some data, reinforcement learning tries to explore the best actions in the medium of operation. Hence, the ability to capture environment through feedback and perform actions makes reinforcement learning suitable for problems involving a series of decisions, i.e. following a policy of actions according to observed environment's state.\\

\begin{figure}
    \centering
    \includegraphics[scale = 0.4]{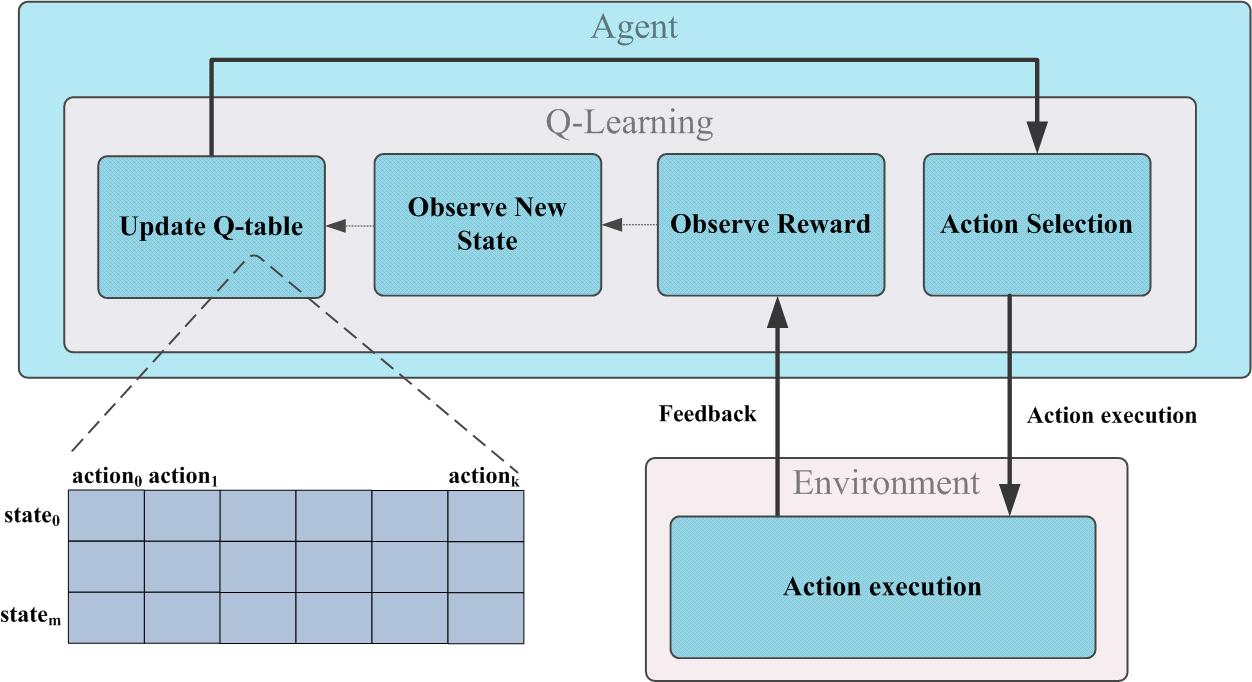}
    \caption{Conceptual diagram of Q-learning operation.}
    \label{fig:QL}
\end{figure}

Reinforcement learning can be model-based or model-free learning. In model-based learning, the agent aims to understand the environment and builds a model to represent it. In contrast, the agent in model-free learning aims to learn a policy to follow. Model-free learning is more suitable for wireless networks since learning from history does not match well with the dynamics of the networks. Q-learning is an example of a model-free reinforcement learning algorithm which aims to learn a policy that tells the agent what action to take under each state. In other words, Q-learning provides the agent the ability to learn the best actions in each state without knowing the model transition probabilities. Fig. \ref{fig:QL} presents a conceptual diagram of Q-learning. The agent starts by selecting an action according to the policy. After action execution, the environment will be influenced and some feedback is returned to the agent. In wireless networks, feedback can be interference or queuing state of nodes or path congestion or learning actions of other nodes and many more factors. Therefore, feedback can either be computed at the agent's side, such as measuring the signal to interference noise ratio. Agents can also exchange their decisions as a form of feedback to each other. The latter poses communication overhead as more network resources must be allocated to facilitate signaling among agents. The feedback will modify the reward and the agent will observe a new state. The state, action, and reward values that are obtained, characterize the quality of the action taken at the current state, hence the agents store this Quality value (i.e., Q-value) in a Q-table. By repeated exploration of different actions in different states, the agent will be able to identify the optimal actions to take. The update of the Q-value is performed iteratively using the Q-learning update equation as follows \cite{Alpaydin}:
\begin{dmath}
	Q(s, a) \gets (1 - \alpha) Q(s, a) + \alpha [R(s, a) + \gamma \max_{a} Q(s, a)],
\label{eq:qvalue}
\end{dmath} 
where $\alpha$ is the learning rate, $\gamma$ is the discount factor, $R(s, a)$ is the reward at state-action pair $(s, a)$ and $Q(s, a)$ is the Q-value of state-action pair $(s, a)$.\\

Besides Q-learning, neural network has been recently used in the state-of-the-art wireless network research. Neural networks are designed to mimic the structure of neurons in the human brain. In particular, a neural network consists of three types of layers: input layer, output layer, and hidden layers. Each layer comprises a set of artificial neurons that perform certain mathematical function, namely neuron activation function. Neurons in a certain layer are connected to the neurons in the preceding layer, where each connection has a weight. In the training phase, the weights are adjusted according to the training dataset. The training dataset provides a set of inputs and the expected outputs. 
\begin{figure}
    \centering
    \includegraphics[scale = 0.45]{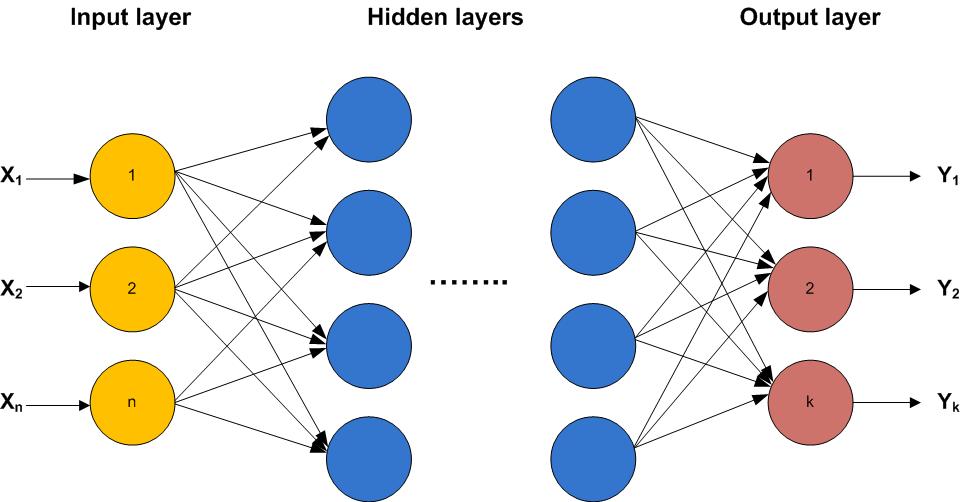}
    \caption{A typical structure of a neural network.}
    \label{fig:NN}
\end{figure}
Neural networks can be structured in different forms such as feedforward, convolutional, or recurrent neural network. A neural network with one hidden layer is a shallow neural network while a neural network with multiple hidden layers is a deep neural network. Furthermore, deep neural network can have different forms such as feedforward, convolutional, recurrent. Fig. \ref{fig:NN} presents an example of a deep feedforward neural network where information flows in one direction. In contrast, a deep recurrent neural network incorporates feedback connections among layers. Such feedback allows the deep recurrent neural network to infer relations in long sequential information (i.e., more efficient at generalization). \\

\begin{figure}
    \centering
    \includegraphics[scale = 0.37]{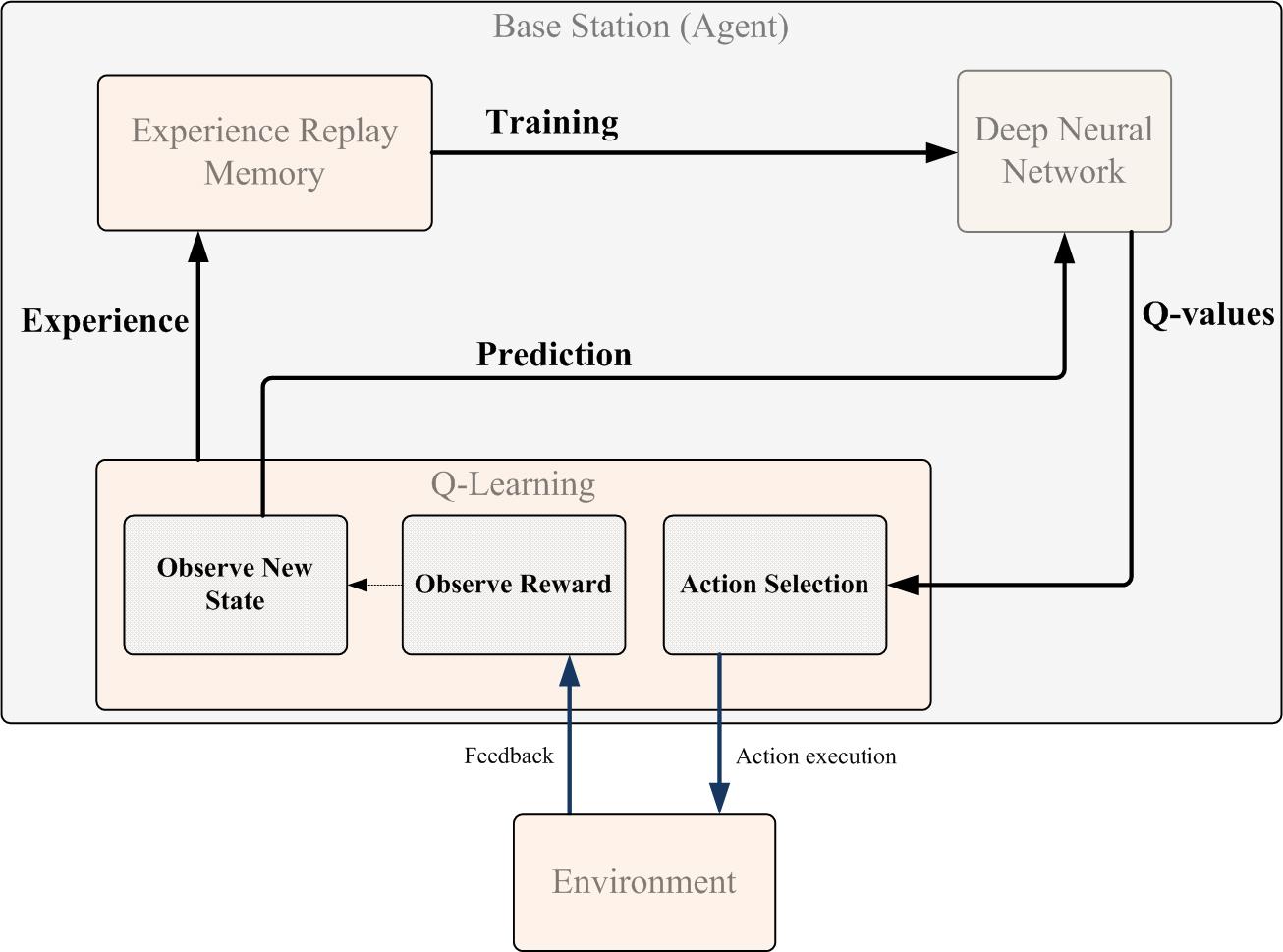}
    \caption{Architecture of Deep Q-learning (Google DeepMind Architecture)}
    \label{fig:DQL_g}
\end{figure}

A more recent method developed by Google DeepMind is the deep Q-learning \cite{DQN}. In contrast to traditional tabular Q-learning, deep Q-learning complements the Q-learning algorithm with a deep convolutional neural network that approximates the Q-value function, avoiding the need to store huge amount of information. Fig. \ref{fig:DQL_g} presents Fig. \ref{fig:DQL_g} presents the Deep Reinforcement Learning components as proposed by Google DeepMind \cite{DQN}. Deep reinforcement learning incorporates Q-learning, a neural network, and an experience replay memory. The Q-learning algorithm is similar to the one presented in Fig. \ref{fig:QL} after removing the Q-table. As such, the Q-learning performs action selection, reward calculation, and observation of new state as before, which is denoted as Q-learning experience \textit{e = {old state (s), old action (a), reward (rc) of old state-action, new state (s$'$)}}. Action selection is performed using Q-learning policy applied to the Q-values estimated by the neural network. The training of the neural network is performed using experiences from the Q-learning algorithm where the training samples are drawn from an experience replay memory that stores experiences over many episodes.\\

The incorporation of deep neural network enabled reinforcement learning to scale to problems that were previously intractable \cite{8103164}. Those problems have high-dimension state-action space that leads to a curse-of-dimensionality problem which might cause slow convergence behavior. The deep neural network acts as a function approximator, where instead of estimating one Q-value each iteration, it predicts the Q-values of the individual actions for a given input state with only a single forward pass. As will be demonstrated later in Section \ref{sec:DMDQ}, this improves convergence of Q-learning. Despite this advantage, tailoring deep neural network to the problem at hand involves the choice of several parameters such as neural network type, number of layers, number of neurons, etc.

\section{Radio Resource Allocation}
\label{sec:RA}

Radio resource allocation is a key function in wireless networks, where efficient utilization of resources is needed for improving the quality of service of the network and quality of experience of the users. Numerous factors impact the performance of radio resource allocation schemes. For instance, wireless channel quality impacts the probability of successful data transmission, hence impacts both throughput and latency. Meanwhile channel quality is affected by interference which is a highly dynamic parameter due to environment, traffic and mobility. In addition, as future wireless networks are anticipated to serve a wider variety of users, spectrum scarcity will be more pronounced and consequently make radio resource allocation more challenging. \\

Radio resource allocation approaches can be classified as centralized or decentralized. Centralized approaches mainly rely on a central entity that collects information from all users in the network. Then, resource allocation is performed according to the global view of the network. On the other hand, decentralized approaches allow users to make decisions autonomously, with slight cooperation in some cases. Centralized approaches can achieve optimal results, however they add overhead of information exchange. Decentralized approaches, on the other hand, allow for more flexibility on the price of sub-optimal results. radio resource allocation schemes also vary according to their optimization objectives such as improving throughput, latency, fairness, energy-efficiency, and spectral-efficiency. Finally, radio resource allocation schemes can be also classified according to the model used, such as optimization-based, heuristic, game-theoretic, or machine learning based. Optimization-based schemes aim for finding the optimal solution to achieve certain objectives. Normally, such problems are complex to solve as it may target optimization of several parameters to achieve multiple objectives simultaneously in the presence of several constraints. Heuristic method is one alternative, where instead of targeting the optimal solution, the method can relax the model assumptions and search for a reasonable solution. However, such relaxation can be loose and the method is not guaranteed to converge to a good solution. Another alternative is game theory-based methods, where network nodes are modeled as players interacting and affecting each others decisions. Each player has a set of actions (decisions) to maximize its payoff (utility). An obvious advantage of game-theoretical methods is its flexibility to adapt to network dynamics. Finally, machine learning schemes are emerging as an alternative solution to the radio resource allocation problem. In the rest of the paper we discuss machine learning methods and their state-of-art usage in wireless networks.

\section{AI-enabled Radio Resource Allocation}
\label{sec:RAML}
In this section, we survey the machine learning-based radio resource allocation techniques and group them based on use cases and application domains.

\subsection{Traditional Mobile Broadband Use Cases}
The research in \cite{8466370} provides an radio resource allocation learner architecture as well as a learning algorithm framework based on reinforcement learning. The architecture is divided into a centralized learner and distributed actors, where the latter is responsible for executing the learned policies and collect samples of experience from the network.\\

In \cite{8425580}, the authors adopt a reinforcement learning approach backed with neural networks to facilitate scheduling decisions such that packet delay and packet drop rate are improved. The problem is formulated as a multi-objective optimization that aims at selecting the optimal scheduling rule and resource block allocation at each transmission time interval. Performance results are conducted for five reinforcement learning algorithms for different windowing factors, different objectives (i.e., delay or packet drop rate) and different traffic classes.\\ 

Radio resource allocation is closely coupled with user association. Users associate to base stations with better channel quality and the number of users associated to a base station impacts the resource sharing decisions. Thus, a collaborative neural Q-learning algorithm is used in \cite{8403664} to perform user-cell association in ultra-dense small-cell networks. Users aim to improve their rate by selecting the best association with base stations. The proposed deep Q-learning algorithm uses the SBSs selected by neighboring users (as the name collaborative implies) as well as its local information as an input to its neural network. The output of the neural network is the estimated Q-values, where each Q-value represents an action (i.e., a SBS to select by the user). 

\subsection{Device-to-Device Communications and Tactile Internet}
Device-to-Device communication is expected to be a significant part of future wireless networks as it provides means to exchange information among users without the need for base station. However, this autonomous operation poses further challenges to resource allocation. Bayesian reinforcement learning is used in \cite{7932486} to perform coalition formation in Device-to-Device networks to maximize network throughput subject to power constraints. Devices form coalitions to maximize their long-term rewards where the decisions used in the coalition formation include the selection of base station, transmission power, transmission channel and transmission mode (e.g., cellular or Device-to-Device). \\

Radio resource allocation also plays a significant role in Tactile Internet which is characterized by its ultra-reliability and ultra-low latency. In \cite{5gforum}, a Q-learning algorithm has been proposed to perform resource block allocations for throughput maximization of data intensive traffic. A two-tier network of small-cell base stations and an eNB is considered to carry the traffic of both conventional and data-intensive users. The Q-learning algorithm learns the traffic patterns and channel conditions of the network, and allocates the resource blocks to users in a way that data-intensive users experience low latency and high throughput. 

\subsection{Unmanned aerial vehicles assisted Networks}
Recently, unmanned aerial vehicles have become a promising approach to augment base stations and thus, enhance the connectivity, capacity and quality of service of wireless networks. The research in \cite{8422503} considers a network of virtual reality users communicating using unmanned aerial vehicles. Unmanned aerial vehicless behave as relays that receive images on the uplink and forwarding them on the downlink using LTE licensed and unlicensed bands, respectively. Furthermore, the quality of images can be adjusted to change the transmitted data size in order to fit the resources allocated. Therefore, dynamic resource allocation is required to both improve the quality of experience of users and to meet the delay requirements of virtual reality applications. To achieve this, the authors employ a Deep Echo-State Network (ESN) algorithm. ESN is a type of recurrent neural network with sparsely connected hidden layers. ESN aims to learn the weights of the output layer only, while weights of hidden layers are fixed and randomly assigned. This in turn increases the speed of learning over conventional recurrent neural networks. 

\section{AI-enabled Deployment, Spectrum Access and Energy Efficiency Techniques} \label{sec:others}
Besides resource allocation, machine learning algorithms are finding significant uses in the deployment of base stations, access to unlicensed bands and energy-efficient network design. In this section, we survey the research works in those domains which is highly relevant to 6G.

\subsection{Deployment of Base Stations in unmanned aerial vehicles-assisted Networks}

Recently, deployment and placement of unmanned aerial vehicles have been active area of research. In \cite{8377340}, the authors address the problem of 3D positioning of aerial base station to assist ground stations for covering mobile users with enhanced quality of service. Due to users' mobility, the network topology gradually changes which impacts the quality of service of users. Furthermore, the positioning algorithm will need more time to re-learn the network. As such, an agile and fast learning algorithm is needed. Authors propose to use a Q-learning algorithm to find an efficient placement of aerial stations to maximize throughput of the network. The difference between current quality of service and previous quality of service (i.e., throughput) constitute the agent's reward. This motivates the agent to improve its decision to gain positive rewards, hence improving the network throughput. After the training phase of Q-learning, it has been shown that the algorithm can adapt to small changes in the network more rapidly.  
 
\subsection{Spectrum Access in Unlicensed Bands}
Deployment of LTE networks in unlicensed bands (i.e., LTE-unlicensed) offer opportunities to improve cellular performance. However, if not properly deployed, LTE-unlicensed may degrade the performance of WLAN, more specifically WiFi. In \cite{8359094}, the authors adopt a proactive resource allocation algorithm to utilize unlicensed spectrum for LTE small cells while maintaining fairness with existing WiFi networks and other LTE operators. In particular, the proposed algorithm aims to balance the spectrum occupancy so as not to degrade the performance of WiFi. To achieve this, the allocation is performed proactively by predicting the LTE traffic and serving it either momentarily or shifting part of it to the future. The proposed approach includes an reinforcement learning with long short-term memory that performs dynamic channel allocation, carrier aggregation, and fractional spectrum access. The use of long short-term memory provides the capability to predict a sequence of future actions, which reinforces the proactive approach.

\subsection{Energy-Efficiency Techniques}
Another important pillar of 6G will be AI-enabled energy efficiency. A reinforcement learning-based, energy-aware resource allocation technique is introduced in \cite{8100645}. In particular, the authors use actor-critic reinforcement learning to find the number of users allocated to each base station, channels and power allocated to each user, as well as selecting the source of energy for the base stations (e.g. energy from the utility grid or renewable energy resources). Due to the stochastic nature of the environment (i.e., wireless channel, renewable energy sources, etc), a model-free learning algorithm is needed. As such, authors propose to use the actor-critic reinforcement learning that combines both policy-based and value-based reinforcement learning. Such integration facilitates learning a continuous action space as well as achieving the convergence to the optimal solution. The actor part is responsible for learning the policy and generate actions, where a Gaussian distribution is used to generate stochastic actions. The critic part evaluates the actor's policy and performs the value function approximation.

\section{Resource Allocation using Deep Q-learning for Low-latency Applications}
\label{sec:DMDQ}

In this section, we present our recent work on utilizing DQL in improving the latency in dense small-cell networks \cite{Globecom}. In this work, mission-critical traffic co-exist with traditional LTE traffic (i.e., non-critical) in a dense and heterogeneous network scenario, in which efficient resource allocation is needed to achieve ultra-low latency for mission-critical nodes. For that purpose, our algorithm, namely Delay Minimization using Deep Q-learning (DMDQ), combines long short-term memory with Q-learning to perform resource block allocation. This study addresses the uplink scheduling problem when critical and non-critical traffic co-exist in a heterogeneous small-cell network where the main objective is to minimize the latency of Mission Critical Devices (MCDs) while maintaining fairness among MCDs and conventional LTE User Equipments (UEs). \\ 

Each base station executes DMDQ in a decentralized manner to minimize the total end-to-end delay of its users where end-to-end delay is defined as transmission and queuing (i.e., scheduling) delays. We define Q-learning tuples as follows: 
\begin{itemize}
    \item \textbf{Agents}: SBSs / eNB
    \item \textbf{States}: We tie Q-learning states to the main objective of delay minimization. Hence, an agent can be either in state 0, in which its average delay is less than a target value, or in state 1 otherwise. 
    \item \textbf{Actions}: Actions are defined as resource block allocations to every user the agent covers. 
    \item \textbf{Reward}: The reward is defined as the Sigmoid function of the total delay. 
    \item \textbf{Policy}: Epsilon-greedy policy is used, where agents either perform exploration by selecting random action, or perform exploitation by selecting the action with maximum Q-value. 
\end{itemize}

We adopt a long short-term memory neural network to estimate Q-values instead of using Q-table method, as discussed in section \ref{sec:ML}.  \\

We compare DMDQ with two algorithms: tabular Q-learning scheme, and a conventional round robin scheme. Our system-level simulator is based on Matlab LTE toolbox. The network consists of a single macro cell network (one eNB) of radius 800m, and a number of Small-cell Base Stations (SBSs) of radius 50m. The number of SBS is 5, the number of UEs is 20, and the number of UNBs (UEs connected to only eNB) is set to 6. MCDs generate traffic according to Beta distribution as defined by 3GPP for machine-type traffic, and UEs generate traffic according to Poisson distribution \cite{Globecom}. \\

\begin{figure}
	\centering
		\includegraphics[scale = 0.25]{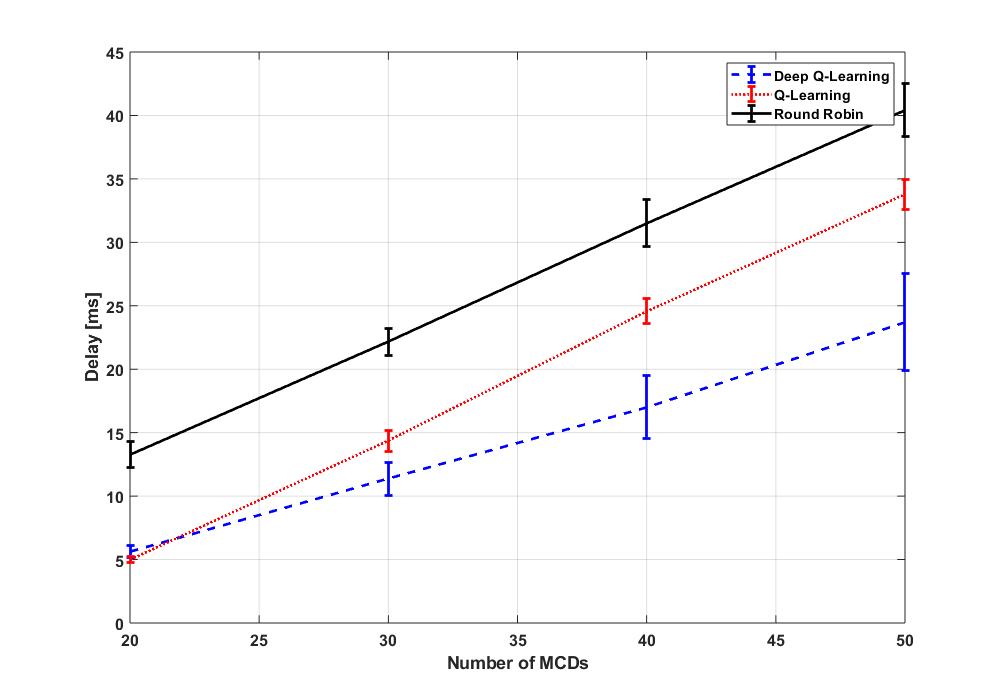}
	\caption{Average end-to-end delay of MCDs [milli-sec].}
	\label{fig:SGD_Delay}
\end{figure}

\begin{figure}
	\centering
		\includegraphics[scale = 0.25]{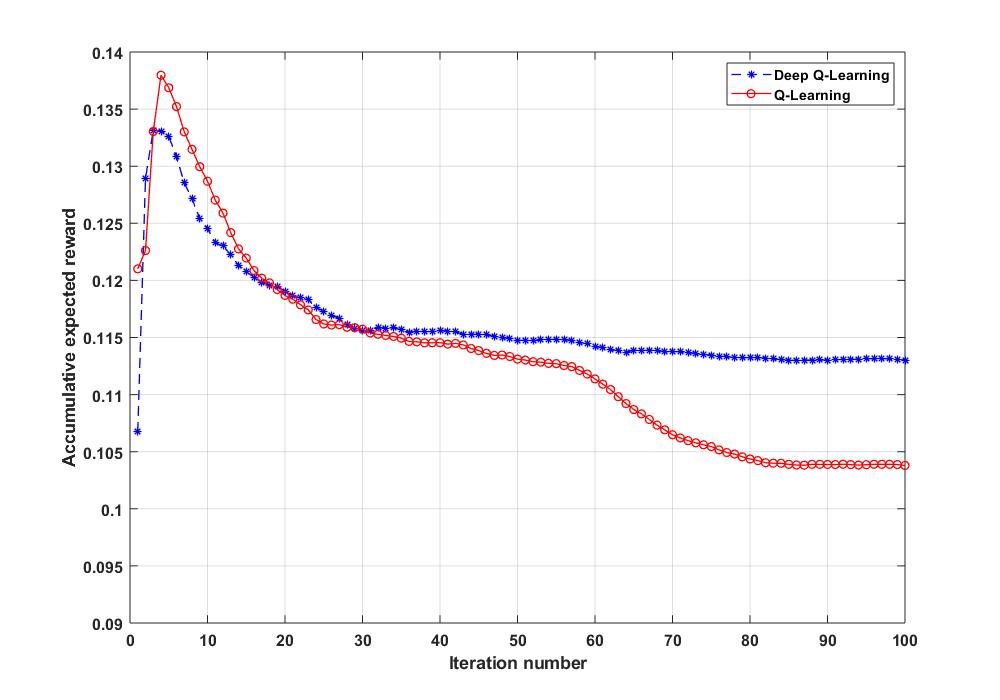}
	\caption{Average discounted reward for both DMDQ and Q-learning.}
	\label{fig:SBS_Conv}
\end{figure}

Fig. \ref{fig:SGD_Delay} shows the average end-to-end delay of all schemes. As observed from the figure, DMDQ outperforms the Q-learning and the round robin scheme by around $28\%$ and $37\%$ delay reduction, respectively. Furthermore, Fig. \ref{fig:SBS_Conv} shows the convergence of DMDQ in comparison to the Q-learning scheme (calculated as $( 1 - \sum\limits_{t = 1}^{T} RC_t / T )$, where $T$ is the subframe number (i.e., x-axis). $\sum\limits_{t = 1}^{T} RC_t / T$ represents the average absolute reward of all SBSs \cite{Globecom}). It can be observed that DMDQ converges more rapidly (i.e., after 30 iterations) while Q-learning requires around 80 iterations to converge. The results show that deep Q-learning algorithm can provide notable reduction in delay with relatively short convergence time. 


\section{Open Issues and Future Directions}
\label{sec:conc}
6G will have a higher level of complexity than all the preceding generations of wireless networks. This will bring in the need for intelligent mechanisms to orchestrate the available resources, services and users. Thus, AI-enabled techniques, or more precisely, the use of machine learning algorithms may allow future networks to learn from their environment, adapt the changes in an automated fashion and achieve optimal performance. \\

Machine learning algorithms have paved the way to significant agility in network management, yet several challenges are still open for research efforts. The open issues can be generally classified into two main pillars: performance of machine learning algorithms and performance of wireless networks.\\

The relatively long convergence time of machine learning methods undermine their usefulness in highly dynamic wireless networks. A careful investigation of the convergence problem and the factors that influence the convergence, are needed. Novel machine learning techniques with faster convergence and online learning capabilities can benefit wireless networks better. \\

Besides convergence, the uncertainty in the wireless network calls for an on-going update of the parameters of the machine learning method or even the method itself. The stochastic nature of the wireless channel may require continuous adaptation. For instance, a network encompassing a large and diverse set of users will have a very dynamic operation. In particular, users who join or leave the network may have very different quality of service and quality of experience requirements. Thus, there is a need to examine whether a ``one size fits all" approach is feasible in real-world implementations.  
In addition, scalability of machine learning algorithms need to be addressed. Machine learning algorithms can become unfeasible for moderately large data, especially in collaborative learning approaches. This calls for a scalable learning algorithm to accommodate for the dense use cases of future wireless networks.\\

Furthermore, supervised and unsupervised learning techniques have been used for massive MIMO recently \cite{Hanzo}. Further research is needed to investigate whether it is possible to enhance the performance of massive MIMO using reinforcement learning and deep learning.\\

Last but not least, AI-enabled networks also impact e-health applications. For instance, advancing outside-of-clinic operations using wearable sensors \cite{Patel2012} requires harmonization of network resource allocation across several technologies and machine learning algorithms can be used for helping with harmonization. Hence, application specific use of machine learning needs to be further explored.

\section*{Acknowledgment}
This research is supported by the Natural Sciences and Engineering Research Council of Canada (NSERC) under RGPIN-2017-03995.

\bibliographystyle{IEEEtran}
\bibliography{reference}

\begin{IEEEbiography}[{\includegraphics[width=0.9in,height=1.3in,clip,keepaspectratio]{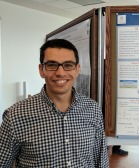}}]{Medhat Elsayed} is a Ph.D. Candidate at the University of Ottawa. He obtained his BSc and MSc degrees from Cairo University, Egypt, in 2009 and 2013 respectively. His research interests are AI-enabled wireless networks, 5G and beyond, and smart grids. 
\end{IEEEbiography}

\begin{IEEEbiography}
[{\includegraphics[width=0.9in,height=1.3in,clip,keepaspectratio]{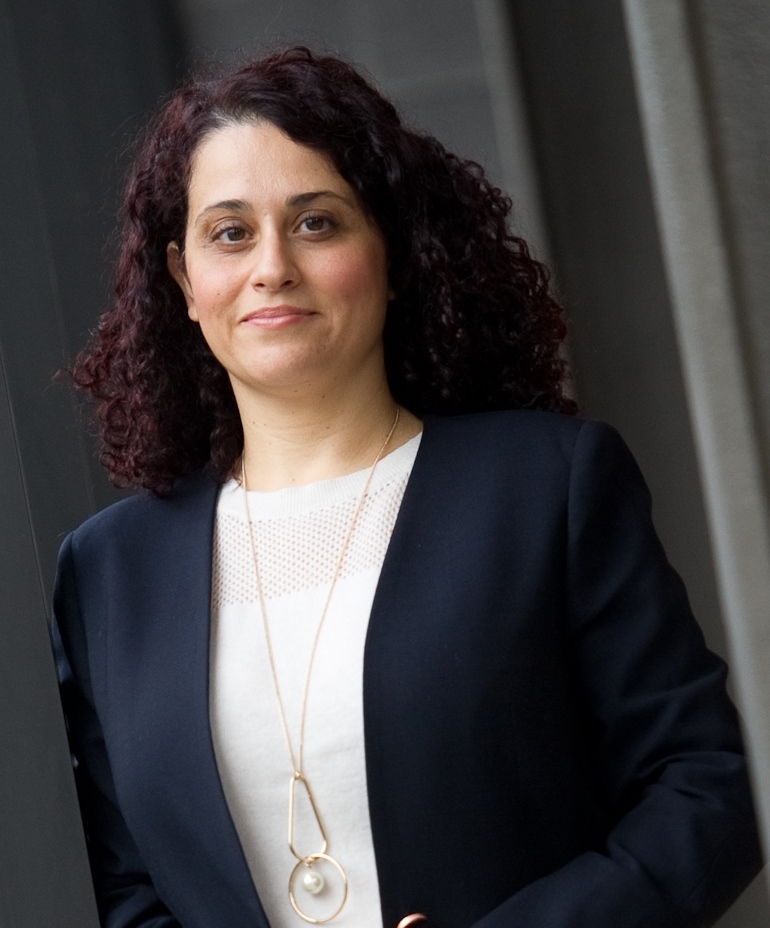}}]{Melike Erol-Kantarci} is an associate professor at the School of Electrical Engineering and Computer Science at the University of Ottawa. She is the founding director of the Networked Systems and Communications Research (NETCORE) laboratory. She is also a courtesy assistant professor at the Department of Electrical and Computer Engineering at Clarkson University, Potsdam, NY, where she was a tenure-track assistant professor prior to joining University of Ottawa. She received her Ph.D. and M.Sc. degrees in Computer Engineering from Istanbul Technical University in 2009 and 2004, respectively. During her Ph.D. studies, she was a Fulbright visiting researcher at the Computer Science Department of the University of California Los Angeles (UCLA). She has over 100 peer-reviewed publications which have been cited over 3900 times and she has an h-index of 30. She has received the IEEE Communication Society Best Tutorial Paper Award and the Best Editor Award of the IEEE Multimedia Communications Technical Committee in 2017. She is the co-editor of two books:  ``Smart Grid: Networking, Data Management, and Business Models'' and ``Transportation and Power Grid in Smart Cities: Communication Networks and Services'' published by CRC Press and Wiley, respectively. She is an editor of the IEEE Communications Letters and IEEE Access. She has acted as the general chair or technical program chair for many international conferences and workshops. She is a senior member of the IEEE and the past vice-chair for Women in Engineering (WIE) at the IEEE Ottawa Section.  She is currently the Chair of Green Smart Grid Communications special interest group of IEEE Technical Committee on Green Communications and Computing. Her main research interests are AI-enabled networks, 5G and beyond wireless networks, smart grid, electric vehicles and Internet of Things.
\end{IEEEbiography}

\end{document}